\documentstyle[11pt,newpasp,twoside,epsf]{article}
\def\edcomment#1{\iffalse\marginpar{\raggedright\sl#1\/}\else\relax\fi}
\reversemarginpar

\begin{document}
\title{Simulation Software: Then, Now and Virtual Observatory}
\author{Peter Teuben}
\affil{Astronomy Department, University of Maryland, College Park, MD 20742, USA}

\begin{abstract}
Like hardware, evolution of software has had a major impact on 
the field of particle simulations.
This paper illustrates how simulation
software has evolved, and where it can go.  In addition, with the
various ongoing Virtual Observatory efforts, producers of data should
think more about sharing their data! Some examples are given of what
we can do with our data and how to share it with our colleagues and observers.
In the Appendix we summarize the findings of an informal data and software
usage survey that we took during this conference.
\end{abstract}


\section{Introduction}

The increased computing speed of both off-the-shelf and
dedicated hardware such as the GRAPE series have made it possible to write
increasingly complex simulation software for very large N-body systems.
What started as simple few
hundred line FORTRAN programs (for a review, see e.g. Aarseth 1999)
with their 
individual analysis routines, have grown into mid-size packages of libraries
and  toolsets. Looking at other fields these
will likely evolve into sophisticated large scale
frameworks such as ROOT (Brun et al. 1999) and 
AIPS++ (Glendenning 1996) if the community combines their programming
efforts and reuse code.

The plan of this paper is as follows. In Section 2 we will show some 
of the current techniques of simulation software, and where this could
lead. In Section 3 the impact of a Virtual Observatory will be discussed,
followed by conclusions and future developments.

\section{Simulation Software}

In the Dark Ages simulation software was developed in small, often
one-person, teams for then available ``supercomputers''. 
Possibly due to limited
electronic communication, software also did not migrate very
easily between researchers. The codes Aarseth developed have
arguably been most widely spread and used (see also Binney \& Tremaine, 1987).
Another early example of shared code development was the OLYMPUS 
programming system, as described in Hockney \& Eastwood (1981).   
These, and upcoming data reduction
packages such as AIPS, GIPSY, IRAF and MIDAS in observational
astronomy have led to a number of packages for particle simulations.
Although these are still 
written by fairly small teams, they 
have now attracted a moderate number of users.
But most notably, they have started to attract 
developers, as is very common in Open Source development these days.
One can distinguish two types of packages:
on the one hand there are
NEMO and Starlab, which present themselves to the user as a collection
of programs that can be called from a shell, or as a collection
of subroutines and functions with which new tools can be built.
On the other hand there are single programs such as {\tt tipsy}
and {\tt astroMD},
which come with their own programmable interface. Our field is, of
course, not that much different from that of for example
High Energy Physics (cf.  De Angelis 2002).


\subsection{NEMO, ZENO}

Initial work on NEMO started in 1986 by Joshua Barnes, Piet Hut and
Peter Teuben (Barnes et al. 1987), and has been subsequently
extended (Teuben 1995). This paper also serves as an update to
document the recent upgrade from Release 2 to 3, of which many
details can be found on the 
NEMO website\footnote{See also {\tt http://www.manybody.org} which
hosts a number of N-body resources}.

\subsubsection{\underline{Source Code}}
The source code consist of two source code trees: 
a ``{\tt src}'' tree and a ``{\tt usr}'' tree, 
which resp. hold the basic NEMO source code, and various public 
(mostly N-body) codes
graciously supplied by their respective authors. 
Most codes in the ``{\tt usr}'' tree are available
``as-is'', some have enhancements for support
within the NEMO environment.
Currently the  ``{\tt usr}'' tree is already about 4 times 
larger than the ``{\tt src}'' tree (860 KLOC vs. 193 KLOC).

Installation in Release 3 has been largely simplified by using
current techniques like {\tt autoconf}, and using a source code
revision control system (CVS) to simplify shared 
development. This has also made it easier to create binary releases.

The source code is largely written in C, with some C++ and FORTRAN
(and support to simplify linking the languages).
Table~\ref{t:codes} lists some of the public code now available
in NEMO.


\begin{table}
\caption{Some of the public N-body codes in NEMO}
\label{t:codes}
\begin{center}
\begin{tabular}{ll}
\tableline
code (author) &
code (author) \\
\tableline

{\tt nbody*} (Aarseth 1999$^{*}$) & {\tt tree++} (Makino)  \\
{\tt ptreecode} (Dubinski)	& {\tt vtc} (Kawaii$^{*}$)  \\
{\tt pmcode} (Klypkin)		& {\tt scfm} (Hernquist \& Ostriker 1998)  \\
{\tt gadget} (Springel)		& {\tt multicode} (Barnes)  \\
{\tt AP3M/hydra} (Couchman)	& {\tt flowcode} (Teuben)  \\
{\tt galaxy} (Sellwood) 	& {\tt yanc} (Dehnen, 2000)  \\
{\tt treecode} (Hernquist 1987) & {\tt superbox} (Richardson, 1999) \\
{\tt treecode1} (Barnes)	& {\tt hackcode1} (Barnes \& Hut, 1986) \\
\tableline
\tableline
\end{tabular}
\end{center}

\footnotesize
$^{(*)}$  see also this volume
\normalsize
\end{table}

\subsubsection{\underline{Packages}}

NEMO's software is packaged and grouped around a number of data formats. In
the {\tt nbody} package, various programs exist to integrate N-body systems with
a wide variety of types of integrators, codes to initialize N-body systems, and
visualization and analysis programs. One of the versatile plotting programs within this
group is called {\tt snapplot}, with which any body variable can be plotted vs.
any other body variable, using on-the-fly code generation (dynamic object
loading) for fast and flexible analysis.
One of its derived programs is {\tt snapgrid},
which produces an image instead of a scatterplot, and includes
effects such as optical depth, and  can
be more directly compared to observations. 
In addition to snapshots and images, two other data formats have a large
set of analysis tools: orbits and tables. Associated with
orbits are potential descriptors, which allow for user supplied potentials
to be loaded into various orbit integrators without the need to recompile
those programs.

In the following example from the nbody/image group of programs
an exponential disk is created, and integrated through
a few dynamical times such that a bar will form. The disk is then
viewed from some angle and a first and second
moment along the new Z axis is
then used to compute a velocity field and velocity dispersion map
on a grid in projected X-Y
space. The resulting dataset can be converted to a FITS file and
manipulated in external packages, such as {\tt saoimage}. The resultant
view of {\tt snapplot} and {\tt ccdplot} is shown 
in Figure 1. 

\footnotesize\begin{verbatim}
% mkexpdisk - 20000 rcut=2 \                  #  make an exponential disk
  | hackcode - - tstop=4 \                    #  integrate to bar formation
  | snaprotate - - 60,45 xz \                 #  rotate around a bit
  | snapgrid - - zvar=-vz moment=-1 times=4 \ #  grid snapshot to velfield
  | tee ngc999vel.ccd \                       #  save a copy of the data 
  | ccdplot - contour=-1:1:0.2 blankval=0     #  contour plot 
% ccdfits ngc9999vel.ccd ngc9999vel.fits      #  convert image to FITS format
% ds9 ngc9999vel.fits &                       #  display with saoimage
\end{verbatim}\normalsize

%


\begin{figure}
\plottwo{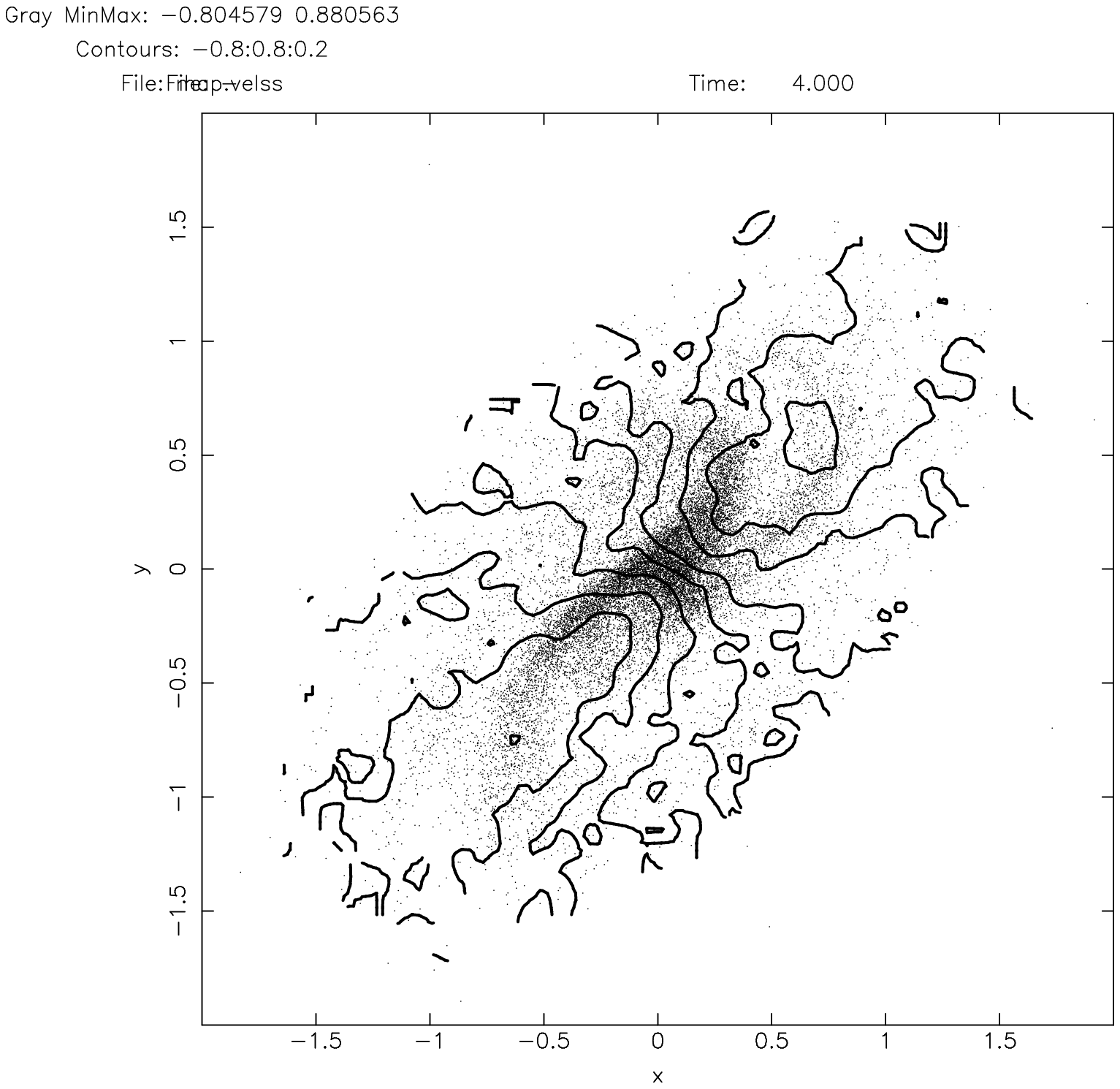}{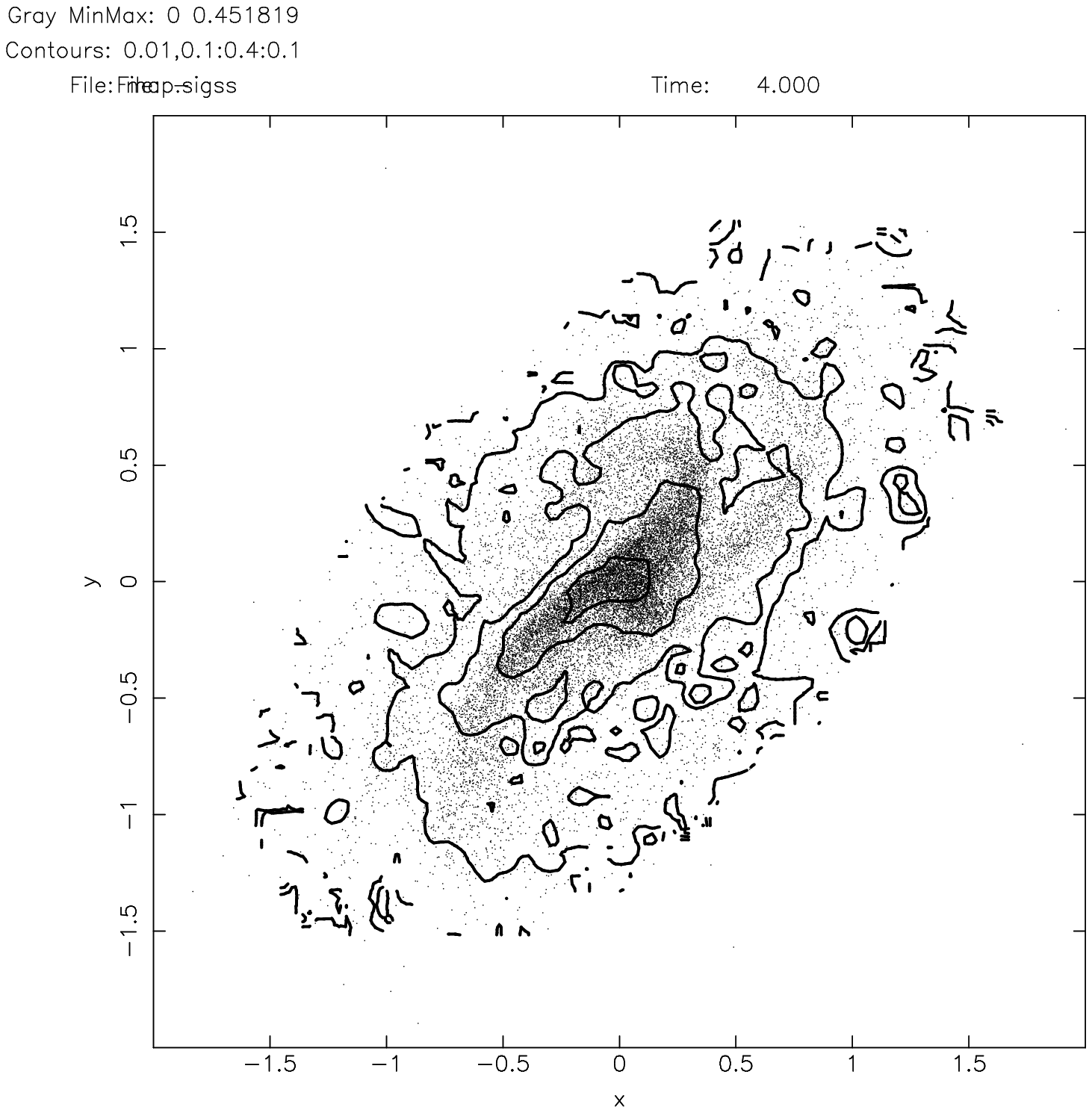}
\caption{Example of graphics output from NEMO. On the left the 
mean velocity (1st intensity weighted moment), on the right
the velocity dispersion (2nd velocity moment, corrected for
mean streaming) in a bar unstable galactic disk. 
}
\label{f:snapplot}
\end{figure}

\subsubsection{\underline{Data Formats}}

NEMO's data format is a structured binary format, where data
elements are identified by name and type, and can be
nested at an arbitrary level. I/O routines access these data
in an associative manner, only retrieving the data
needed at that time. Data is also interchangeable between
machines with different data types.

With the introduction of more  ``foreign''
integrators into NEMO, each with their own data format, it became 
necessary to be able to read and write a large variety of data
formats. Because there
is no such thing as a standard interchange format like FITS
\footnote{N-body data can often be described as a table, and thus
the FITS {\tt BINTABLE} format could very well serve as an
an interchange format, see e.g. Teuben 1995}, NEMO's {\it snapshot}
format has become the central format to interchange format X to Y.

\subsubsection{\underline{ZENO}}

The ZENO software package is an evolutionary product of an earlier
version of NEMO, written by Joshua Barnes, and is largely still source
code compatible with NEMO. ZENO instead concentrates on N-body and SPH
simulations, and particle representations are dynamically
extendible instead.

\subsection{Starlab}

Starlab (Portegies Zwart et al. 2001) was loosely modeled after NEMO,
but written completely from scratch to handle the more intricate
physics of collisional dynamics. A new tree-based data structure was
introduced to handle the more complex stellar interactions. In
addition, data piped through the system would not lose information
not known to the data-handler. 
The code is
mostly written in C++, and as of this writing consists of about 236
KLOC in 911 files.
The {\tt kira} integrator (23 KLOC) can also be linked with a variety of
GRAPE libraries to take advantage if this hardware is available.
{\tt kira} also contains the {\tt SeBa} stellar evolution module.


Here is a simple example how to create a King model with a given IMF,
binaries and stellar evolution, then integrated on a GRAPE-6 and output
dumped in a file that can be processed using a number
of tools available in Starlab. 

\footnotesize\begin{verbatim}
% makeking -n 500 -w 5 -i -u \                # 500 particle king model
   | makemass -f 1 -x -2.0 -l 0.1 -u 20 \     # mass spectrum
   | makesecondary -f 0.1 -l 0.1 \            # make secondary stars
   | add_star -Q 0.5 -R 5 \                   # add stellar evolution
   | scale -M 1 -E -0.25 -Q 0.5 \             # scale to virial equilibrium
   | makebinary -f 1 -l 1 -u 1000 -o 2 \      # set orbitals for binaries
   | kira_grape6 -t 100 -d 1 -D 10 -f 0.3 -n 10 -q 0.5 -G 2 -S -B  > big.out

\end{verbatim}\normalsize

Visualization in Starlab is currently best done with
{\tt partiview}, which can take advantage of the hierarchical
space-time nature of the data and also knows about double stars
(Teuben et al. 2001).

\subsection{Scripting}

In the tradition of UNIX, the modular design of NEMO and Starlab
(and their common general data format) programs 
can be combined in a large variety of
ways. Prototyping can now be done
in minutes, to produce complex and sophisticated
analysis pipelines. However, we have found in practice that
the resulting shell scripts (sh, csh, make) are often not very 
robust. 
Subsequently scripting has exhibited some fragility that more comprehensive
control mechanisms could help prevent.
Future versions or new software should make use of such control data in a more
reliable way. Another approach is to use an embeddable scripting language, such
as {\tt python} or {\tt ruby}, which can enforce a tighter connection 
between codes and data. In addition, these hybrid software environments often
lend themselves to better GUI development. We are currently experimenting
with this.

\subsection{TIPSY}

Another popular package to analyze particle simulations is 
TIPSY\footnote{See also {\tt http://www-hpcc.astro.washington.edu/tools/tipsy/tipsy.html}}
(Quinn \& Katz).
This package has followed the philosophy of
a single program, with a special command interpreter to operate
on snapshots containing a combination of SPH,
dark-matter, and pure Newtonian particles.


\footnotesize\begin{verbatim}
  % tipsy
  <yes, Peter>openascii run99.ascii
  <yes, Peter>readascii run99.bin
  read time 14.970800
  <yes, Peter>xall
  <yes, Peter>quit
  <I will miss you, master>
  % tipsysnap run99.ascii - | snapplot - xrange=-4:4 yrange=-4:4
\end{verbatim}\normalsize

The obvious advantage of this approach is speed, as the data always
remains in memory. However, any modifications to the program means
it will have to be abandoned in order to be recompiled. Plug-ins
or dynamic objects can alleviate some of these problems.

\subsection{IDL, and other}

As the field of astrophysical particle simulations is a very
specialized one, the majority
of research is done with personal codes, and likewise their analysis.
In recent years these codes have also incorporated more
interesting physics, such as 
simple empirical sticky particle dynamics,
SPH gas dynamics, stellar evolution, chemodynamical evolution, etc.
These codes are mostly written in languages like FORTRAN, C or C++.
In recent years commercial graphics packages like
IDL and Open Source toolkits such as VTK (Visualization ToolKit) 
have also become popular to analyze and visualize such
complex datasets. 
Generic visualization packages such as AVS, 
IRIS Explorer and IBM's Data Visualizer are widely used,
yet limitations in such generic packages continue to create
a niche for programs like
the recently developed AstroMD toolkit (Becciani et al. 2000).
This program uses the VTK 
library to allow for very
sophisticated multi-variate data analysis and visualization.


\section{Virtual Observatory}

The concept of a Virtual Observatory to federate various
observational databases (e.g.~Brunner et al.~2001) and 
make combined searching and analysis on these databases possible, has not
gone un-noticed in the theoretical community. Teuben et al. 2002
argued that adding various types of theoretical data to the VO will
benefit observers as well as theoreticians, and open up new and
unexplored avenues for research.

For example, existence of a standard number of
(benchmark) datasets will benefit authors of new codes
to quickly compare and highlight differences between
various codes. This is of course not something new,
but still a relatively rare event in our community.
After the first published code comparison by Lecar (1968) 
it still took nearly 30 years for the community
to continue this effort when in 1997, Heggie (2002, this volume)
reported on a comparison between different star cluster
simulations and Sellwood (1997) 
published a comparison between five different 
N-body codes typically used in galaxy simulations.
Setting up test problems is important (Heggie 1997, 2001), and 
has been a standard in many field of computational science
(e.g. Stone \& Norman 1992).

In a Virtual Observatory we can expect to select models, compare them
to existing data and perform various types of fits to best describe
the observations. In addition, new models can be compared to old
models, and provide feedback to code development. 
%
%
%
%
In order to better understand the scope of the role of theory in a
Virtual Observatory, we have started to construct a 
``toymodel''\footnote{See {\tt http://www.astro.umd.edu/nemo/tvo}},
which contains a growing collection of different types of theory data.
The only condition for data to be added to
this toymodel is that they must either
be benchmark data, or datasets associated with published papers.

%
%
%
%
%
%
%
%
%
%
%
%
%

\section{Conclusion}

We have reviewed the evolution of particle simulation codes,
and seen them adapt to the ever growing speeds of hardware.
This will have to include the rapid development of
PC cluster hardware, which will require enhancing
the scalability of parallel algorithms.
The simulation software itself has more slowly matured
(e.g. compare different software engineering practices)
with promises of code reuse and extendibility. 
The future in software development is likely going to be in
frameworks such as ROOT and AIPS++ with plenty of room for
niche applications, as long as they can easily share their
data!

A Virtual Observatory framework will allow for a more seamless
integration of observations and simulations, and allow
astronomers to compare observations and obtain
best fit models. It should also enable theory to develop
more in pace with simulations and encourage code reuse and data format
sharing.


It is expected that the N-body simulation community will continue
to contribute, as in the past, a diversity of insights that will
continue to make the field exciting and productive for years to
come.

\acknowledgments
I acknowledge support from the Alfred P. Sloan foundation, through a
grant to Piet Hut for research at the Hayden Planetarium of the
American Museum of Natural History in New York.  I also wish to thank
the AAS International Travel Grant for support; 
Joshua Barnes and Piet Hut for our collaborations in NEMO; the
numerous users of NEMO in (requests for) enhancing the software; Vicky Johnson
for a careful reading; 
George Lake and Tom Quinn for long discussions around the future of
particle pushing analysis frameworks, a.k.a. the N-Chilada.

\section*{Appendix: Data Usage Survey}

Before the final ``Open Discussion'' (chaired by Prof. Sugimoto) 
a small survey was handed out to all participants in order get some more
insight in the types of data that are produced in our simulations, and
the current habits of its practitioners. 
Although no head-count was made of the number of people present
at this session, 53 (exactly 50\% of the 106 
officially registered participants)
forms were returned and analyzed.

A similar (but unpublished) survey was held in 1994
amongst a much smaller but similar focus group of astronomers
at the Lake Tahoe meeting to celebrate Sverre Aarseth's 60th birthday.
The current "survey" was intended to aid the discussion on typical 
current "N-body" data usage and future Virtual 
Observatories. A quick overview of the results was
given during the summary session on Friday afternoon. Here we reproduce
some of that discussion.

\subsection*{The Survey}

\footnotesize\begin{verbatim}
1. Your name or alias:

2. Do you:  (in all of the items below you are allowed to 
	    mark multiple items, perhaps you can indicate 
	    the percentage of relevance in each)
      a  develop your own code(s)	[name(s): 
      b	 use existing code(s)		[name(s):
      c  other 
      d  N/A

3. What is your typical range in N:

4. Do you
      a	 save particle/grid data (see also 5)
      b	 discard data (e.g. analysis within simulation code)
      c	 other:
      d  N/A

5. Your data format is mostly:
      a	 table of particle attributes (m, x, y, z, vx, ..)
      b	 grid of cell attributes (den(i,j,k), vx(i,j,k), ..)
      c	 hydrid of 1 and 2 (e.g. P3M)
      d  a tree structure
      e  other: (try and describe)
      f  N/A

6. What analysis software do you use:
      a  my own			[name(s):    / language(s):
      b	 sm
      c	 IDL
      d	 NEMO
      e	 ZENO
      f	 Starlab
      g	 Tipsy
      h	 other:
      i  N/A

7. What kind of ancillary data do you also store:
      a	 minor diagnostics (E, Lx, Ly, Lz, ...)
      b	 neighrest neighbor list
      c  detailed grouping info
      d	 a tree structure representing particles
      e	 a tree structure representing grid
      f	 other:

8. Do you use models
      a  to compare to your own other models
      b	 to compare to other people models
      c	 to compare/fit to observational data
      d	 other:
      e  N/A

9. If you do any of those, do you
      a  compare in configuration (pos, vel) space (e.g. velocity field)
      b  compare in derived quantities  (e.g. power spectrum, luminosity function)
      c	 other:
      d  N/A

10.Any suggestions for Virtual Observatories?
      a  a waste of time, because ....
      b  a great idea, but ....
      c  other:
      d  N/A

11.Do you have any data yourself you have available?
      a  no
      b  yes, but I can't make it available
      c  yes, and I might be able to make it available
      d  yes, they are available on the web already

12. Any remaining comments?

\end{verbatim}\normalsize

\subsection*{Results}

It quickly became clear that making a good survey is harder than it
looks. Many questions had multiple answers, thus the numbers quotes
will be percentages and will generally add up to over 100\%.

\begin{enumerate}
\item[{\bf 1.}]
Nobody was using an alias, 
everybody choose to use their known name (one person choose his/her Japanese name).

\item[{\bf 2.}]
a) 72\% b) 68\% c) 4\% d) 2\newline
Codes mentioned (quite a few respondents actually missed 
the request to mention names of their code(s)): 
AMR using boxlib,
ap3m, 
asph,
c++tree, 
chameleon,
COSMIC, 
eurostar, 
gadget, 
gasoline, 
gizmo, 
grapesph, 
hydra, 
kira, 
nemo, 
nbodyN, 
p3m, 
p3msph, 
pg, 
pkdgrav, 
pmtreecode, 
ptreecode, 
scf, 
superbox.
tipsy, 
treescf, 
treesph.

\item[{\bf 3.}]
2 - 1,000,000,000, with one respondent quoting $\le 32768$!

\item[{\bf 4.}]
a) 90\% b) 6\% c) - d) - \newline
data summaries, expansion coefficients, test vector.

\item[{\bf 5.}]
a) 84\% b) 14\% c) 14\% d) 14\% e) 6\% f) 4\% \newline
series of nested adaptive grids, distribution function, 

\item[{\bf 6.}]
a) 60\% b) 36\% c) 32\% d) 10\% e) - f) 10\% g) 20\% h) - i) :\newline
xmgr(2), gnuplot(8), mongo(4), dx(2), amrvis(1), pgplot(1),
pdl (1), midas(1), sm(1)
\newline
Languages: Fortran (8), C (6), C++ (4), Perl (2), Awk (1)

\item[{\bf 7.}]
a) 74\% b) 22\% c) 12\% d) 12\% e) - f) - \newline
special purpose diagnostic files, likelyhoods, never store data.

\item[{\bf 8.}]
a) 74\% b) 66\% c) 70\% d) - e) 4\%

\item[{\bf 9.}]
a) 64\% b) 64\% c) - d) 4\%

\item[{\bf 10.}]
{\it ambitious, good luck, hope it takes off, would not use it } (2x)

\item[{\bf 11.}]
a) 18\% b) 6\% c) 50\% e) 16\%

\item[{\bf 12.}]
A nice variety of comments:
{\it The VO might be a good place to
offer their codes to the public. There needs to be a common
data format (CDF, HDF). The project is very ambitious.
Worry that the credit to a persons work is lost. VO should
be looked at as an internationally funded observatory.
Theory should also play their parts in making images
available for public outreach.  
Programming resources are needed to support the scientists
who are supposed to this work. Reliability and possible
refereeing needed for the submitted data. Good luck!
Useful to chain simulations: the end of one simulation
can be used as input to the next.}

\end{enumerate}

\subsection*{Discussion and Conclusions}

Although a very large programming effort is shared amongst this
community, at least an equal amount is using simulation software
from colleagues (and probably expanding it).
The survey unfortunately
did not address the question how data is interchanged between codes,
if any, despite that a good fraction of people use more than one code.

Data analysis fills an almost equally wide spectrum, from a variety
of special purpose software (TIPSY, NEMO) to utilizing
generic, and sometimes even commercial, software (IDL, sm, perl, dx).
Part of this is no doubt sociological and 
the familiarity of the user with that specific software.

Although a large amount of software is still written in FORTRAN, 
the majority is now in C/C++. Perhaps notable in this survey was the
absence of Java.
{\tt perl} and {\tt awk} are popular scripting languages,
although nobody mentioned the basic shells {\tt sh} and {\tt csh}. Also,
no explicit operating systems were mentioned.
The large amounts of code, and interesting genealogy between them,
shows that most of them fill a specific niche in the market. 

\end{document}